\documentstyle[preprint,aps,epsfig,axodraw]{revtex}

\begin{document}

\draft

\preprint{
\hfill$\vcenter{ \hbox{\bf MADPH-99-1136}
                {\hbox{\bf IFT-P.063/99}
                 \hbox{\bf UH-511-936}
                 \hbox{\bf August 1999}}
}$ }

\title{Diphoton Signals for Large Extra Dimensions \\
at the Tevatron  and CERN LHC}
 
\author{O. J. P. \'Eboli$^1$, T.\ Han$^2$, M.\ B.\ Magro$^2$, and P. G. 
Mercadante $^3$}

\address{$^1$Instituto de F\'{\i}sica Te\'orica, Universidade
Estadual  Paulista \\
Rua Pamplona 145, S\~ao Paulo, SP 01405--900, Brazil.}

\address{$^2$Department of Physics, University of Wisconsin \\
Madison, WI 53706, USA.}

\address{$^3$Department of Physics and Astronomy, University of Hawaii \\
Honolulu, HI 96822, USA}

\vskip -0.75cm 

\maketitle
\begin{abstract}
  
\vskip-5ex 

  We analyze the potentiality of hadron colliders to search for large extra
  dimensions via the production of photon pairs. The virtual exchange of
  Kaluza--Klein gravitons can significantly enhance this processes provided
  the quantum gravity scale ($M_S$) is in the TeV range.  We studied in detail
  the subprocesses $q \bar{q} \to \gamma \gamma$ and $g g \to \gamma \gamma$
  taking into account the complete Standard Model and graviton contributions
  as well as the unitarity constraints.  We show that the Tevatron Run II will
  be able to probe $M_S$ up to 1.5--1.9 TeV at 2$\sigma$ level, while the LHC
  can extend this search to 5.3--6.7 TeV, depending on the number of extra
  dimensions.

\end{abstract}



\section{Introduction}
 
The known constructions of a consistent quantum gravity theory require the
existence of extra dimensions \cite{gsw}, which should have been
compactified. Recently there has been a great interest in the possibility that
the scale of quantum gravity is of the order of the electroweak scale
\cite{led} instead of the Planck scale $M_{pl} \simeq 10^{19}$ GeV. A simple
argument based on the Gauss' law in arbitrary dimensions shows that the Planck
scale is related to the radius of compactification ($R$) of the $n$ extra
dimensions by
\begin{equation}
                M^2_{pl} \sim R^n M_S^{n+2} \; ,
\end{equation}
where $M_S$ is the $(4+n)-$dimensional fundamental Planck scale or the string
scale. Thus the largeness of the $4-$dimensional Planck scale $M_{pl}$ (or
smallness of the Newton's constant) can be attributed to the existence of
large extra dimensions of volume $R^n$. If one identifies $M_S \sim {\cal
O}$(1 TeV), this scenario resolves the original gauge hierarchy problem
between the weak scale and the fundamental Planck scale, and lead to rich low
energy phenomenology \cite{led,ph-astro}.  The $n=1$ case corresponds to
$R\simeq 10^8$ km, which is ruled out by observation on planetary motion. In
the case of two extra dimensions, the gravitational force is modified on the
$0.1$ mm scale; a region not subject to direct experimental searches yet.
However, astrophysics constraints from supernovae has set a limit $M_S > 30\
\hbox{TeV}$ for $n=2$ \cite{astro}, and thus disfavoring a solution to the
gauge hierarchy problem as well as direct collider searches.  We will
therefore pay attention to $n \ge 3$.

Generally speaking, when the extra dimensions get compactified, the fields
propagating there give rise to towers of Kaluza-Klein states \cite{KK},
separated in mass by ${\cal O}(1/R)$.  In order to evade strong constraints on
theories with large extra dimensions from electroweak precision measurements,
the Standard Model (SM) fields are assumed to live on a 4--dimensional
hypersurface, and only gravity propagates in the extra dimensions. This
assumption is based on new developments in string theory
\cite{string0,string,scs}.  If gravity becomes strong at the TeV scale,
Kaluza--Klein (KK) gravitons should play a r\^ole in high--energy particle
collisions, either being radiated or as a virtual exchange state. There has
been much work in the recent literature to explore the collider consequences
of the KK gravitons
\cite{ph-astro,joanne,giudice,mathews,tao_grav,pheno,kingman}.  In this work
we study the potentiality of hadron colliders to probe extra dimensions
through the rather clean and easy-to-reconstruct process
\begin{equation}
	p p~(\bar{p}) \to \gamma \gamma X\; ,
\end{equation}
where the virtual graviton exchange modifies the angular and energy
dependence of the $ q \bar{q}~ (gg) \to \gamma \gamma$ processes. In
the Standard Model the gluon fusion process takes place via loop
diagrams. However, it does play an important part in the low scale
gravity searches since its interference with the graviton exchange is
roughly proportional to $M_S^{-4}$ instead of $M_s^{-8}$ 
for pure Kaluza--Klein exchanges. Moreover, we expect this
process to be enhanced at the LHC due to its large 
gluon--gluon luminosity.

Since the introduction of new spin--2 particles modifies the high
energy behavior of the theory, we study the constraints on the
subprocess center--of--mass energy for a given $M_S$ in order to
respect partial wave unitarity. Keeping these bounds in mind, we show
that the Tevatron Run II is able to probe $M_S$ up to 1.5--1.9 TeV,
while the LHC can extend this search to 5.3--6.7 TeV, depending on the
number of extra dimensions.  We also study some characteristic
kinematical distributions for the signal.

In Sec.~\ref{section2} we evaluate the relevant processes for the
diphoton production in hadron colliders. We then examine the
perturbative unitarity constraints for the $\gamma \gamma \to \gamma
\gamma$ process in Sec.~\ref{unit}. Our results for the Tevatron are
presented in Sec.~\ref{tevatron}, while Sec.~\ref{lhc} contains
results for the LHC. We draw our conclusions in Sec.~\ref{section4}.

\section{The Diphoton Production}
\label{section2}

\begin{figure}
\vskip 36pt
\begin{center}
\centerline{
\begin{picture}(400,80)(-15,-15)
\Gluon(10,20)(50,20){-4}{4}
\Gluon(10,60)(50,60){-4}{4}
\ArrowLine(50,20)(50,60)
\ArrowLine(50,60)(90,60)
\ArrowLine(90,60)(90,20)
\ArrowLine(90,20)(50,20)
\Photon(90,20)(130,20){4}{3}
\Photon(90,60)(130,60){4}{3}
\Text(5,20)[r]{$g$}
\Text(5,60)[r]{$g$}
\Text(45,40)[r]{$q$}
\Text(70,70)[c]{$q$}
\Text(95,40)[l]{$q$}
\Text(70,10)[c]{$q$}
\Text(135,20)[l]{$\gamma (k')$}
\Text(135,60)[l]{$\gamma (k)$}
\Gluon(200,10)(240,40){-4}{4}
\Gluon(200,70)(240,40){4}{4}
\Photon(240,40)(280,40){4}{5}
\Photon(240,40)(280,40){-4}{5}
\Photon(280,40)(320,10){4}{3}
\Photon(280,40)(320,70){-4}{3}
\Text(195,10)[r]{$g$}
\Text(195,70)[r]{$g$}
\Text(260,50)[c]{$KK$}
\Text(325,10)[l]{$\gamma (k')$}
\Text(325,70)[l]{$\gamma (k)$}
\end{picture}
}
\end{center}
\vskip -0.8cm
\caption{Feynman diagrams contributing to the subprocess 
$gg \to \gamma\gamma$, including the Kaluza--Klein graviton exchange.
The crossed diagrams are not displayed.}
\label{graph_gg}
\end{figure}
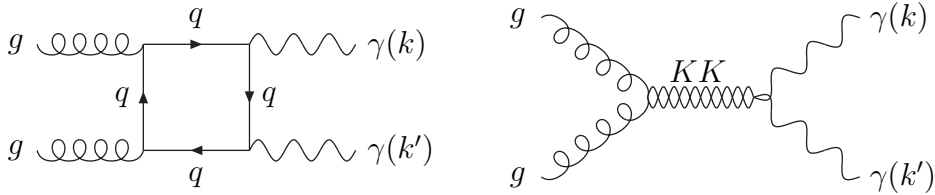

In hadronic collisions, photon pairs can be produced by quark--antiquark
annihilation 
\begin{eqnarray}
q\bar{q} \to \gamma\gamma \; ,
\label{signalq}
\end{eqnarray}
as well as by gluon--gluon fusion
\begin{eqnarray}
g g \to \gamma\gamma \; . 
\label{signalg}
\end{eqnarray}
The lowest order diagrams contributing to these processes are
displayed in Figs.~\ref{graph_qq} and \ref{graph_gg} respectively.  We
only consider the contribution from spin-2 KK gravitons and neglect
the KK scalar exchange since it couples to the trace of the
energy--momentum tensor and consequently is proportional to the mass
of the particles entering the reaction.

The differential cross section of the process (\ref{signalq}),
including the SM $t$-- and $u$--channel diagrams and the exchange KK
gravitons in the $s$--channel, is given by
\begin{eqnarray}
\frac{d\sigma (q\bar{q}\to \gamma\gamma)}{d\cos\theta} &=& 
\frac{1}{192\pi\hat{s}} \left[ 32\alpha^2 \pi^2 
\frac{1+\cos^2\theta}{1-\cos^2\theta} + 
\frac{\kappa^4 |D(\hat{s})|^2}{256} \hat{s}^4 \sin^2\theta~ (1+\cos^2\theta )
\right. \nonumber \\
&& - \left.{\rm Re}\left(i\pi\alpha \kappa^2 D^{\ast}(\hat{s}) ~\hat{s}^2~ 
(1+\cos^2\theta )\right)\right] \; ,
\label{dsigma_qq}
\end{eqnarray}
with $\hat{s}$ being the center--of--mass energy of the subprocess and
$\alpha$ the electromagnetic coupling constant. 
$\kappa$ is the gravitational coupling
while $D(\hat{s})$ stands for the sum of graviton propagators
of a mass $m_{KK}^{}$:
\begin{equation}
	D(\xi) = \sum_{KK} \frac{1}{\xi - m_{KK}^2} \;\; .
\label{dfunction}
\end{equation}
Since the mass separation of $1/R$ between the KK modes is rather
small, we can approximate the sum by an integral, which leads to
\cite{tao_grav}
\begin{equation}
\kappa^2 D(\hat{s}) \to 16\pi~ \hat{s}^{-1 + n/2}~ M_S^{-(n+2)}~
 \left[ \pi + 
2i I\left(\frac{\Lambda}{\sqrt{\hat{s}}} \right)\right]\; ,
\label{propagator}
\end{equation}
where $n$ is the number of extra dimensions and the non-resonant 
contribution is
\begin{equation}
I\left(\frac{\Lambda}{\sqrt{\hat{s}}} \right) = 
P\int^{\Lambda/\sqrt{\hat{s}}}_0 dy \frac{y^{n-1}}{1-y^2}\; .
\label{I}
\end{equation}
Here we have introduced an explicit ultraviolet cutoff $\Lambda$.
Na\"{\i}vely, we would expect that the cutoff is of the order of the
string scale. We thus express it as
\begin{equation}
  \Lambda = r M_S \; ,
\label{ratio}
\end{equation}
with $r\sim {\cal O}(1)$. Taking $\Lambda=M_S \gg \sqrt{\hat{s}}$, we
have for Eq.~(\ref{propagator})
\begin{equation}
\kappa^2 D(\hat{s}) \to -\frac{32i\pi}{(n-2)M_S^4}\;\;\;,\;\;{\rm for}\;\; 
n>2 \; . 
\label{propagator2}
\end{equation}
It is interesting to note that in this limit, the power for $M_S$ is
independent of the number of extra dimensions. Our results in this
limit agree with the ones in Ref.~\cite{kingman}. However, we would
like to stress that it is important to keep the full expression
(\ref{propagator}) since there is a significant contribution coming
from higher $\sqrt{\hat{s}}$.

In the Standard Model, the subprocess (\ref{signalg})
takes place through 1-loop quark-box 
diagrams in leading order; see Fig.~\ref{graph_gg}. 
The differential cross section for this subprocess is then
\begin{equation}
\frac{d\sigma(gg\to\gamma\gamma )}{d\cos\theta } = \frac{1}{64\pi\hat{s}} 
\cdot \frac{1}{4} \cdot \frac{1}{8} \cdot \frac{1}{8}~
\left| \sum_{a;b;\lambda_{1,2,3,4}} ({\cal M}_{grav}^{\lambda_1
\lambda_2\lambda_3\lambda_4} + {\cal M}_{sm}^{\lambda_1\lambda_2
\lambda_3\lambda_4})\right|^2 \;  .
\label{dsigma_gg}
\end{equation}
We present the one--loop SM helicity amplitudes ${\cal
M}_{sm}^{\lambda_1\lambda_2\lambda_3\lambda_4}$ in an appendix, which
agree with the existing results in the literature \cite{old:gg}. As
for the helicity amplitudes ${\cal M}_{grav}^{\lambda_1 \lambda_2
\lambda_3 \lambda_4}$ for the $s-$channel exchange of massive spin--2
KK states, they are given by
\begin{eqnarray}
&&{\cal M}^{+-+-}_{grav} = {\cal M}^{-+-+}_{grav} = - \delta_{ab}~
\frac{i}{16}~\kappa^2~ 
D(\hat{s})~ \hat{s}^2~(1+\cos\theta)^2 \; , \label{pmpm_grav} \\
&&{\cal M}^{+--+}_{grav} = {\cal M}^{-++-}_{grav} = - \delta_{ab}~
\frac{i}{16}~\kappa^2~ 
D(\hat{s})~ \hat{s}^2~(1-\cos\theta)^2 \; , \label{pmmp_grav}
\end{eqnarray}
where the remaining amplitudes vanish and $a$ and $b$ are the color of the
incoming gluons. 

As we can see from Eqs.\ (\ref{dsigma_qq}) and (\ref{dsigma_gg}), the
cross sections for diphoton production grow with the subprocess
center--of-mass energy, and therefore violate unitarity at high
energies. This is the signal that gravity should become strongly
interacting at these energies or the possible existence of new states
needed to restore unitarity perturbatively. In order not to violate
unitarity in our perturbative calculation we excluded from our
analyses very high $\gamma \gamma$ center--of--mass energies, as
explained in the next section.

\section{Unitarity bound from $\gamma \gamma \to \gamma \gamma$}
\label{unit}

We now examine the bounds from partial wave unitarity on the ratio
$M_S/\sqrt{s}$ and $r$ in Eq.~(\ref{ratio}).  For simplicity, we
consider the elastic process $\gamma \gamma \to \gamma\gamma$. The
$J-$partial wave amplitude can be obtained from the elastic matrix
element ${\cal M}$ through
\begin{equation}
a^J_{\mu;\mu^\prime} = \frac{1}{64 \pi} 
\int^1_{-1} d\cos\theta\; d^J_{\mu;\mu^\prime} \left[ 
-i{\cal M}^{\lambda_1\lambda_2\lambda_3\lambda_4}\right] \; ,
\label{aj}
\end{equation}
where $\mu =\lambda_1-\lambda_2$, $\mu^\prime = \lambda_3-\lambda_4$,
and the Wigner functions $d^J_{\mu;\mu^\prime}$ follow the conventions
of Ref.\ \cite{pdg}. Unitarity leads to
\begin{equation}
	\hbox{Im}~ a^J_{\mu;\mu}  \le | a^J_{\mu;\mu} |^2 \; ,
\end{equation}
which implies that 
\begin{equation}
	|\hbox{Re}~ a^J_{\mu;\mu}|  \le \frac{1}{2} \; .
\label{uni1} 
\end{equation}
After the diagonalization of $a^J_{\mu;\mu^\prime}$, the largest
eigenvalue ($\chi$) has modulus $|\chi| \le 1$. The critical values
$|\chi|= 1$ and $|{\rm Re}(\chi )| = 1/2$ define the approximate limit
of validity of the perturbation expansion. In our analyses we verified
that the requirement of $|\chi |\le 1$ leads to stronger bounds for
almost all values of $r$.
 
At high energies the $\gamma\gamma$ elastic scattering amplitude is
dominated by the KK exchange, and consequently we neglected the SM
contribution. Taking into account the graviton exchange in the $s$,
$t$, and $u$ channels we obtain that
\begin{eqnarray}
{\cal M}^{++++} &=& {\cal M}^{----} = -\frac{i}{4}~s^2 \kappa^2~\left[D(t)+
D(u)\right]\; ; \label{m_gggg1}\\
{\cal M}^{+-+-} &=& {\cal M}^{-+-+} = -\frac{i}{4}~u^2 \kappa^2~\left[D(s)+
D(t)\right]\; ; \label{m_gggg2}\\
{\cal M}^{+--+} &=& {\cal M}^{-++-} = -\frac{i}{4}~t^2 \kappa^2~\left[D(s)+
D(u)\right]\; , \label{m_gggg3}
\end{eqnarray}
where the $D(s)$, $D(t)$, and $D(u)$ stand for the graviton propagator
in Eq.~(\ref{dfunction}) in the $s$, $t$, and $u$ channels
respectively; their explicit expressions after summing over the KK
modes can be found in Ref.\ \cite{tao_grav}\footnote{Here we rectify
the expression for $D(t)$ ($D(u)$), for $n$ odd, which should be given
by
\[
D(t) = \frac{|t|^{n/2-1}}{\Gamma (n/2)}\frac{R^n}{(4\pi )^{n/2}} (-2i) 
I_E(M_S/\sqrt{|t|})\; ,
\]
with
\[
I_E(M_S/\sqrt{|t|}) = (-1)^{(n-1)/2}\left[ \sum_{k=1}^{(n-1)/2} \frac{(-1)^k}
{2k-1}\left( \frac{M_S}{\sqrt{|t|}}\right)^{2k-1} + \tan^{-1}(M_S/\sqrt{|t|})
\right]\; .
\]}.  
Notice that the imaginary (real) part of ${\cal M}$ 
comes from the (non-)resonant sum over the
KK states; see Eq.\ (\ref{propagator}).

Bose--Einstein statistics implies that only the even $J$ partial waves are
present in the elastic $\gamma \gamma$ scattering.  From the expressions
(\ref{m_gggg1})$-$(\ref{m_gggg3}) we obtain that the non-vanishing $J=0$ and 2
independent partial waves for n=2 are
\begin{eqnarray}
	a^0_{0;0} &=& \frac{is^2}{4M_S^4} \left[x^2\ln 
\left(\frac{1}{x^2}+1\right)+\ln\left(x^2+1\right)\right]\; ,
\\
	a^2_{0;0} &=& \frac{is^2}{8M_S^4}x^2\left[\left(4x^4+3x^2+2\right)
\ln\left(1+\frac{1}{x^2}\right)-4x^2-4\right]\; ,
\\
	a^2_{2;2} &=& \frac{-s^2}{40M_S^4}\left[\pi-i\ln\left(x^2-1\right)
\right]+\frac{is^2}{640M_S^4}\left[x^2\left(80+160x^2+160x^4+80x^6+
16x^8\right)\times\right. \nonumber\\
&&\times\ln\left(\frac{1}{x^2}+1\right)+\left.16\ln\left(1+x^2\right)-
2x^2\left(\frac{154}{3}+\frac{188}{3}x^2+36x^4+8x^6\right)\right]\; ,
\\
	a^2_{2;-2} &=&  a^0_{2;2}\; ,
\end{eqnarray}
and for $n=3$ 
\begin{eqnarray}
	a^0_{0;0} &=& \frac{is^2\Lambda}{2M_S^5} - \frac{is^{5/2}}{6M_S^5} 
\left[ 2\tan^{-1}(x) + x - x^3 \ln\left(1+\frac{1}{x^2}\right)\right]\;,
\\
	a^2_{0;0} &=& -\frac{is^2}{14M_S^5}\Lambda\left[ 2 - \frac{3}{10}
(14+10x^2) + \frac{70+252x^2+180x^4}{30} - \frac{2}{15x}\tan^{-1}(x) - 
\right. \nonumber \\
&&\left. -\frac{70x^2+252x^4+180x^6}{30}\ln\left(1+\frac{1}{x^2}\right)
\right]\;, 
\\
	a^2_{2;2} &=& -\frac{s^{5/2}}{40M_S^5}\left\{ \pi + 2i\left[ -x + 
\frac{1}{2}\ln\left(\frac{\Lambda + \sqrt{s}}{\Lambda - \sqrt{s}}\right)
\right]\right\} + \frac{is^2}{20M_S^5}\Lambda - \nonumber \\
&& -\frac{is^2}{576M_S^5}\Lambda\left[ 2 - \frac{4}{21}(72+14x^2) - 
\frac{1848+2304x^2+1300x^4+280x^6}{35} + \right.\nonumber \\
&&\left. + \frac{2048}{35x}\tan^{-1}(x) + \frac{2520+3360x^2+3024x^4+1440x^6+
280x^8}{35}\ln\left(1+\frac{1}{x^2}\right)\right]\;,
\\
	a^2_{2;-2} &=&  a^0_{2;2}\; ,
\end{eqnarray}
with $x = r M_S/\sqrt{s}$ and by parity $a^2_{2;2} = a^2_{-2;-2}$ and
$a^2_{-2;2} = a^2_{2;-2}$.  Since the matrix $a^2_{i,j}$ has a very simple
form, it is easy to obtain that its non-vanishing eigenvalues $\chi_i$ are
	$a^2_{0;0} \;\;\;\; \hbox{and} \;\;\;\; 2~a^2_{2;2}$. 

\begin{figure}[p]
\begin{center}
\psfig{file=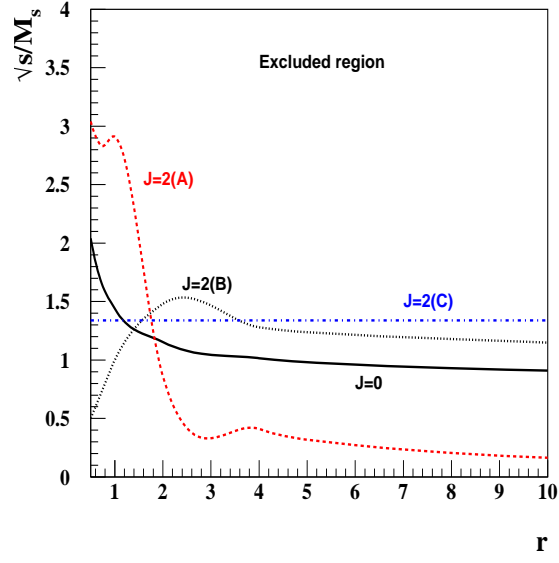,height=8cm,width=8cm}
\end{center}
\caption{$J=0$ and 2 unitarity bounds in the plane $\protect\sqrt{s}/M_S
\otimes r$ for the elastic $\gamma\gamma$ scattering and $n=2$. $J=2$(A) 
corresponds to the limit from the eigenvalue $\chi = |a^2_{0;0}| < 1$, (B) to
$2|a^2_{2;2}| < 1$ and (C) to $2|Re(a^2_{2;2})| < 1/2$.}
\label{fig:unit2}
\end{figure}

\begin{figure}
\begin{center}
\psfig{file=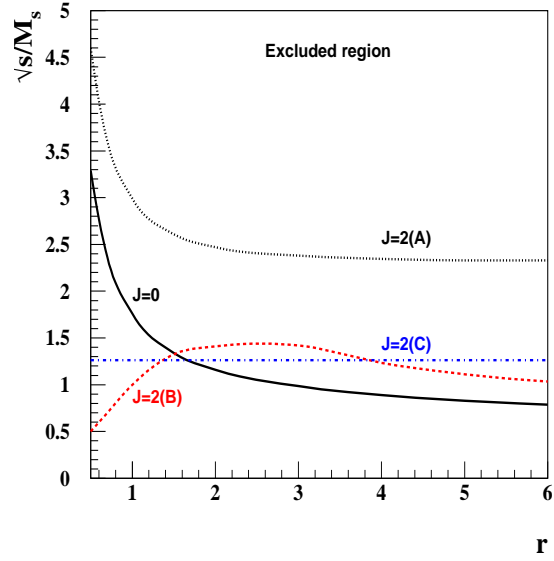,height=8cm,width=8cm}
\end{center}
\caption{Same as in Fig.\ \protect\ref{fig:unit2} but for $n=3$.} 
\label{fig:unit3}
\end{figure}

We exhibit in Figs.~\ref{fig:unit2} and \ref{fig:unit3} the excluded region in
the plane $\sqrt{s}/M_S \otimes r$ stemming from the requirement that
$|\chi_i| \le 1$ and $|{\rm Re}(\chi_i)| \le 1/2$ for $n=2$ and 3.  As we can
see from Fig.\ \ref{fig:unit2}, the $J=2$ partial wave leads to more stringent
bounds than the $J=0$ wave for almost all values of $r$ for $n=2$. Meanwhile,
in the case of $n=3$, the $J=0$ partial wave leads to stronger bounds for
large $r$. Assuming that the ultraviolet cutoff $r=1$, we find that
perturbation theory is valid for $\sqrt{\hat{s}} < M_S$.  This is what one may
naively anticipated. Asymptotically for larger $r$, the limit approaches to
$\sqrt{\hat{s}} \le 0.7M_S$ for $n=3$, and $\sqrt{\hat{s}} \le 0.1M_S$ for
$n=2$.  In the rest of our calculations we make the conservative choice $r=1$
and $\sqrt{\hat{s}} \le 0.9 M_S$.

\section{Numerical Studies at Hadron Colliders}
\subsection{Results for Tevatron}
\label{tevatron}

Due to the large available center--of--mass energy, the Fermilab
Tevatron is a promising facility to look for effects from low scale
quantum gravity through diphoton production in hadronic collisions. At
Run I, both CDF and D\O\ studied the production of diphotons
\cite{cdf-d0}. The CDF probed higher $\gamma\gamma$ invariant masses
in the range $50 < M_{\gamma\gamma} < 350$ GeV, but the binning
information on the data is not available. It is nevertheless possible
to exclude $M_S \le 0.91$ (0.87) TeV for $n=3$ (4) at 95\% CL
\cite{kingman}.

For Run II there will be a substantial increase in luminosity (2
fb$^{-1}$) as well as a slightly higher center--of--mass energy (2
TeV). Therefore, we expect to obtain tighter bounds than the ones
coming from the Run I. We evaluated the diphoton production cross
section imposing that
\begin{itemize}
	\item $|\eta_\gamma | <1$;

	\item $p_T^\gamma > 12$ GeV;
	
	\item $350 \hbox{ GeV }< M_{ \gamma\gamma} < 0.9~ M_S$.
\end{itemize}
Notice that we introduced a upper bound on $M_{\gamma\gamma}$ in order
to guarantee that our perturbative calculation does not violate
partial wave unitarity; see Sec.\ \ref{unit}. The strong cut on the
minimal $M_{\gamma\gamma}$ reduces the background from jets faking
photons \cite{cdf-d0} to a negligible level, leaving only the
irreducible SM background. After imposing those cuts and assuming a
detection efficiency of 80\% for each photon, we anticipate 12
reconstructed background events at Run II per experiment.

\begin{figure}
\begin{center}
\psfig{file=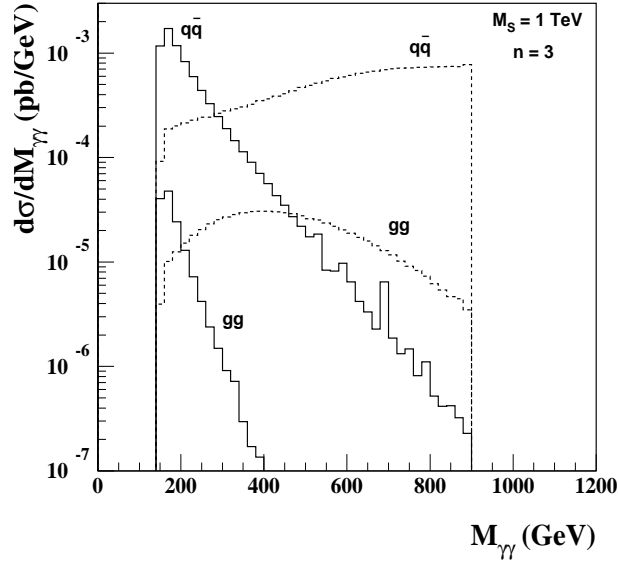,height=8cm,width=8cm}
\end{center}
\caption{$\gamma\gamma$ spectrum originated from the SM (solid lines) and
graviton--exchange (dashed) contributions for $q \bar{q}$ and gluon--gluon
fusions at the Tevatron Run II. In this figure we took $M_S = 1$ TeV,
$n=3$, and applied the cuts described in the text, using now
$M_{\gamma\gamma} > 150$ GeV.}
\label{fig:qqgg}
\end{figure}

We display in Fig.\ \ref{fig:qqgg} the $M_{\gamma\gamma}$ spectrum
from the SM and graviton--exchange contributions for $q
\bar{q}$ and gluon--gluon fusions. We use in our calculation the MRSG
distribution functions \cite{mrsg} and took the $M_{\gamma\gamma}$ as the QCD
scales. As we can see from this figure, the $q \bar{q}$ fusion
dominates both the SM and graviton contributions and  the graviton
exchange is the most important source of diphoton at large
$\gamma\gamma$ invariant masses.  This behavior lead us to introduce
the above minimum $M_{\gamma\gamma}$ cut to enhance the signal.

We obtained the $2\sigma$ limits for the quantum gravity scale $M_S$ 
requiring that
\begin{equation}
\frac{\sigma_{tot} - \sigma_{sm}}{\sqrt{\sigma_{sm}}}\sqrt{{\cal L}} 
\;\epsilon \ge 2 \; ,
\label{limit}
\end{equation}
where $\epsilon = 0.80$ is the reconstruction efficiency for one photon,
$\sigma_{sm}$ is the SM cross section, and $\sigma_{tot}$ the total cross
section including the new physics; see Sec.\ \ref{section2}. We present in
Table \ref{limtab1} the $2\sigma$ attainable bounds on $M_S$ at Run II for
several choices of $n$ and an integrated luminosity of ${\cal L} = 2$
fb$^{-1}$, corresponding to the observation of 19 events for SM plus
signal. Notice that the bounds are better than the ones for Run I
\cite{kingman} because the higher luminosity allows us to perform a more
stringent cut on the minimum $M_{\gamma\gamma}$, consequently enhancing the
graviton exchange contribution. Furthermore, these bounds do not scale as
$(n-2) M_S^4$ as expected from Eq.\ (\ref{propagator2}), showing the
importance to take into account the full expression for the graviton
propagator.

\begin{figure}
\begin{center}
\psfig{file=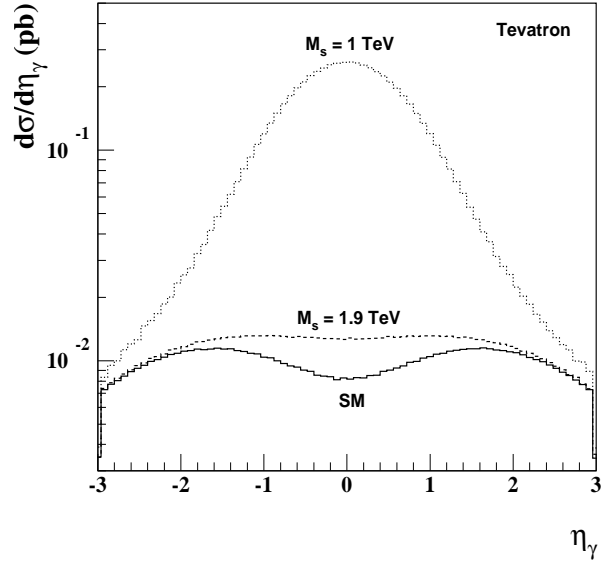,height=8cm,width=8cm}
\end{center}
\caption{Rapidity distributions of the photons for the SM background (solid
line) and for the signal for $M_S = 1.9$ TeV (dashed line) and $M_S = 1$ TeV
(dotted line) with $n = 3$ at the Tevatron Run II. We imposed the cuts
$|\eta_{\gamma}| < 3$, $350$ GeV $< M_{\gamma\gamma} < 0.9 M_S$ and
$p_{T}^{\gamma} > 12$ GeV.}
\label{figtev1}
\end{figure}

\begin{figure}
\begin{center}
\psfig{file=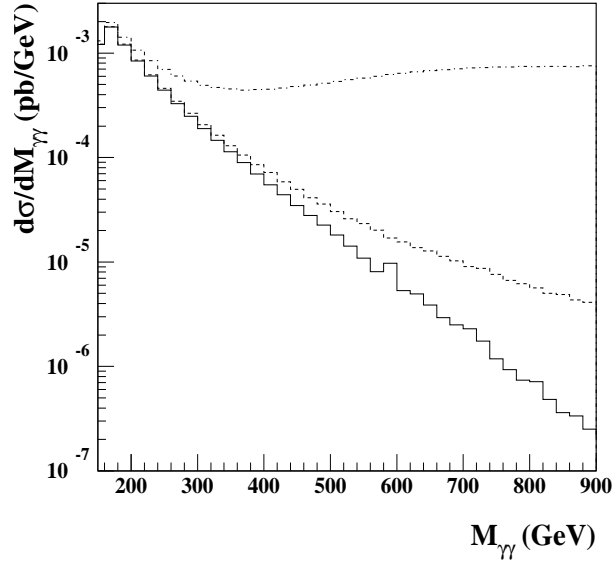,height=8cm,width=8cm}
\end{center}
\caption{Invariant mass distributions of the photon pair for the SM 
background (solid line) and for the signal for $M_S = 1.9$ TeV (dashed 
line) and $M_S = 1$ TeV (dot-dashed line) with $n = 3$ at the Tevatron Run II.
We imposed the cuts $|\eta_{\gamma}| < 1$, $150$ GeV $< M_{\gamma\gamma} 
< 0.9 M_S$ and $p_{T}^{\gamma} > 12$ GeV.}
\label{fig:minv_tev}
\end{figure}

In order to establish further evidence of the signal from low scale
quantum gravity after a deviation from the SM is observed, we study
several kinematical distributions, comparing the predictions of the SM
and the quantum graviton exchange. We show in Fig.\ \ref{figtev1} the
photon rapidity distributions after the invariant mass and transverse 
momentum cuts, relaxing the rapidity to be $|\eta_\gamma | < 3$, for the SM 
and for the $n = 3$ gravity
signal with $M_S = 1.9$ TeV, which corresponds to the $2\sigma$ limit,
and $M_S = 1$ TeV, which leads to a larger anomalous contribution. As
we expected, the $s$--channel KK exchange gives rise to more events at
low rapidities in comparison with the $t$ and $u$--channel SM
background, indicating that an angular distribution study could be
crucial to separate the signal from the background. As remarked above,
the rapid growth of the $M_{\gamma\gamma}$ spectrum is an important
feature for KK graviton exchange; see Fig.\ \ref{fig:minv_tev}.


\subsection{Results for LHC}
\label{lhc}

The CERN Large Hadron Collider (LHC) will be able to considerably
extend the search for low--energy quantum gravity due to its large
center--of--mass energy ($\sqrt{s}=14$ TeV) and high luminosity
(${\cal L} = 10$--$100$ fb$^{-1}$). Therefore, we have also analyzed
the diphoton production at LHC in order to access its potentiality to
probe $M_S$ via this reaction.

\begin{figure}
\begin{center}
\psfig{file=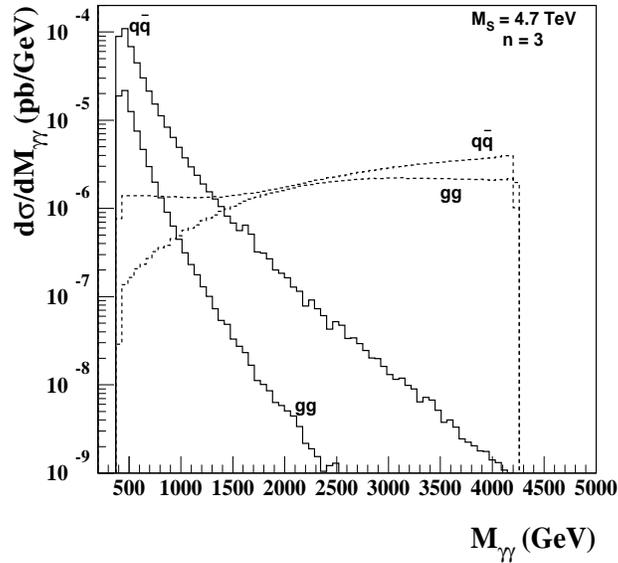,height=8cm,width=8cm}
\end{center}
\caption{$\gamma\gamma$ spectrum originated from the SM (solid lines) and
graviton--exchange (dashed) contributions for $q \bar{q}$ and gluon--gluon
fusions at the LHC. We took $M_S = 4.7$ TeV, $n=3$, and applied the
cuts described in the text, using now $M_{\gamma\gamma} > 400$ GeV.}
\label{lhc_parts1}
\end{figure}

At the LHC there will be a very large gluon--gluon luminosity which
will enhance the importance of $g g \to \gamma \gamma$ subprocess, as can 
be seen from Fig.\ \ref{lhc_parts1}.
Therefore, we can anticipate that the interference between the SM loop
contribution to this process and the KK graviton exchange will play an
important r\^ole in determining the limits for $M_S$ since it is comparable 
to the leading $q\bar{q}$ anomalous contribution for large $M_S$. 

At high $\gamma\gamma$ center--of--mass energies the main background
is the irreducible SM diphoton production. In order to suppress this
background and enhance the signal we applied the following cuts:
\begin{itemize}

	\item we required the photons to be produced in the central region of
	the detector, {\em i.e.} the polar angle of the photons in the
	laboratory should satisfy $| \cos \theta_\gamma| < 0.8$;

	\item $M_{\gamma \gamma} > 0.8~(1)$ TeV, according with the
	luminosity ${\cal L} = 10~(100)\;{\rm pb}^{-1}$, in order to
	use the fast decrease of the SM background with the increase
	of the subprocess center--of--mass energy;

	\item $M_{\gamma \gamma} < 0.9~ M_S$ to be sure that partial wave
	unitarity is not violated.

\end{itemize}
The SM contribution after cuts has a cross section of 1.86 (0.83) fb, which
corresponds to 18 (83) reconstructed events for ${\cal L} = 10~(100)$ 
fb$^{-1}$.

In Table \ref{limtab2}, we present the $2 \sigma$ attainable limits on $M_S$
for different number of extra dimensions and two integrated luminosities
${\cal L} = 10$ fb$^{-1}$ and $100$ fb$^{-1}$, corresponding to the
observation of 27 and 101 events for the SM plus signal respectively. In our
calculation we applied the above cuts and used the MRSG parton distribution
functions with the renormalization and factorization scales taken to be
$M_{\gamma\gamma}$.  As we expected, the LHC will be able to improve the
Tevatron bounds by a factor $\simeq 4$ due to its higher energy and
luminosities. Moreover, our results are slightly better than the other limits
presented so far in the literature \cite{joanne,giudice,mathews}.

\begin{figure}
\begin{center}
\psfig{file=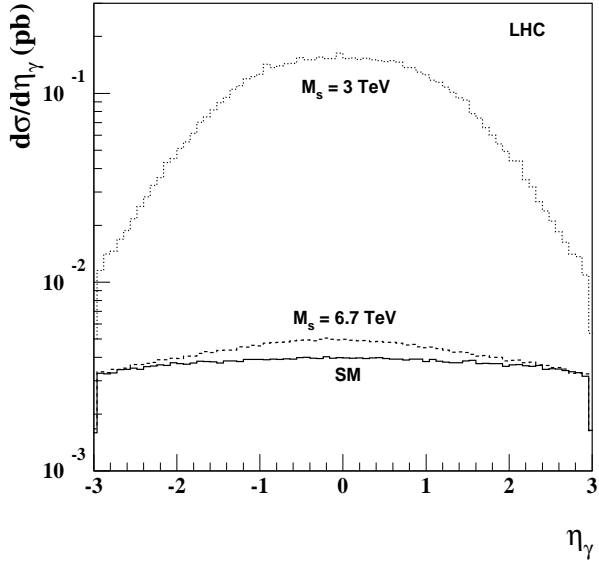,height=8cm,width=8cm}
\end{center}
\caption{Rapidity distributions of the photons for the SM background 
(solid line) and for the signal for $M_S = 6.7$ TeV (dashed line) and
$M_S = 3$ TeV (dotted line) with $n = 3$ at the LHC. We imposed
the cuts $|\eta_\gamma | < 3$ and $800$ GeV $< M_{\gamma\gamma} < 0.9 M_S$.}
\label{figlhc1}
\end{figure}

\begin{figure}
\begin{center}
\psfig{file=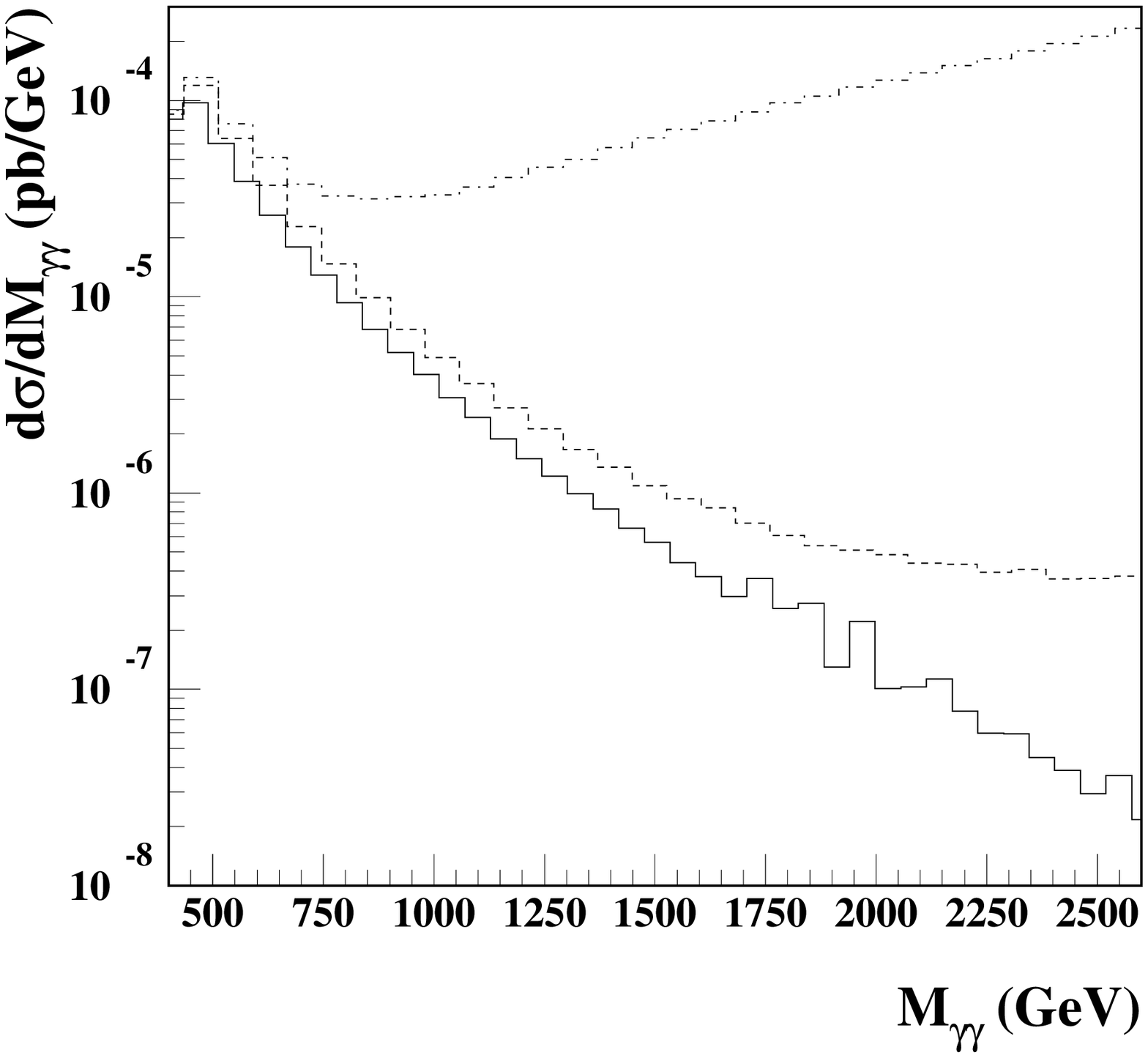,height=8cm,width=8cm}
\end{center}
\caption{Invariant mass distributions of the photon pair for the SM
 background (solid line) and the signal for $M_S = 6.7$ TeV (dashed
 line) and $M_S = 3$ TeV (dot--dashed line) with $n = 3$ at the LHC.
 We imposed the angular cut $| \cos \theta_\gamma| < 0.8$ and $400$ GeV
 $< M_{\gamma\gamma} < 0.9 M_S$.}
\label{figlhc2}
\end{figure}

In order to learn more about the KK states giving rise to the signal,
we should also study characteristic kinematical distributions. For
instance, the rapidity distributions due to graviton exchange are
distinct from the SM one since the signal contribution takes place via
$s$--channel exchanges while the backgrounds are $u$-- and
$t$--channel processes.  We show in Fig.\ \ref{figlhc1} the photon
rapidity spectrum after the invariant mass cut and requiring that 
$|\eta_\gamma | < 3$, 
including the signal for $M_S = 6.7$ TeV
and $M_S = 3$ TeV with $n = 3$ and the SM backgrounds. As expected,
the distribution for the graviton signal is more central than the
background.  Analogously to the Tevatron analysis, the KK modes also
show themselves in the high diphoton invariant mass region as seen in
Fig.\ \ref{figlhc2}.

\section{Discussion and Conclusions}
\label{section4}

High center--of--mass energies at hadron colliders provide a good opportunity
to probe the physics with low--scale quantum gravity.  We have analyzed the
potentiality of hadron colliders to search for large extra dimensions signals
via the production of photon pairs. Although the virtual exchange of
Kaluza--Klein gravitons may significantly enhance the rate for this processes
and presents characteristically different kinematical distributions from the
SM process, it is sensitive to an unknown ultraviolet cutoff $\Lambda$, which
should be at ${\cal O}(M_S)$. We examined the constraints on the relation
between $\sqrt s$ and $M_S$ from the partial wave unitarity as a function of
this cutoff. Keeping in mind the unitarity constraint, we calculated in detail
the subprocesses $q \bar{q} \to \gamma \gamma$ and $g g \to \gamma \gamma$
taking into account the complete Standard Model and graviton contributions. We
found that the Tevatron is able to probe $M_S$ at the 2$\sigma$ level up to
1.5--1.9 TeV at Run II, for 7--3 extra dimensions; while the LHC can extend
this search to 5.3 (6.5)--6.7 (8.5) TeV for a luminosity ${\cal L} = 10$
$(100)$ fb$^{-1}$.

It is important to notice that our results are better, or at least 
comparable, to the ones presented in the literature so far. At the Tevatron, 
it is already clear that the Run II will provide stronger limits on the 
new graviton scale due simply to the enhancement of the luminosity. Our 
analysis shows that a careful kinematical study for the diphoton production 
would improve the limits obtained so far in the Run II from other process as 
Drell-Yan 
production \cite{joanne}, $p\bar{p}\rightarrow \hbox{jet} + E\!\!\!\!/\;$ 
\cite{giudice} and top production \cite{mathews}. In the LHC, besides the 
kinematical cuts suggested here, the inclusion of the SM box diagrams for 
$gg\rightarrow\gamma\gamma$ also improves the signal, since its contribution 
in the 
interference level is no longer negligible. As a matter of fact, our results 
are comparable to the best ones presented so far, obtained from the process 
$p\bar{p}\rightarrow \hbox{jet} + E\!\!\!\!/\;$ by Giudice et. al. 
\cite{giudice}. 

\acknowledgments

This research was supported in part by the University of Wisconsin
Research Committee with funds granted by the Wisconsin Alumni Research
Foundation, by the U.S.\ Department of Energy under grant
DE-FG02-95ER40896 and DE-FG03-94ER40833, by Conselho Nacional de
Desenvolvimento Cient\'{\i}fico e Tecnol\'ogico (CNPq), by
Funda\c{c}\~ao de Amparo \`a Pesquisa do Estado de S\~ao Paulo
(FAPESP), and by Programa de Apoio a N\'ucleos de Excel\^encia
(PRONEX).

\vskip 18pt

\noindent{\bf Note added:} When we were preparing this manuscript we became
aware of Ref.\ \cite{outros} which also studies the diphoton production at the
LHC.


\appendix

\section{Helicity Amplitudes for the Box Diagrams}\label{app_a}

The independent helicity amplitudes for the SM process $gg \to
\gamma\gamma$ including only the contribution of a massless quark are
\begin{eqnarray}
{\cal M}^{++++}_{sm}(s,t,u) &=& -i~ \delta_{ab}~
 8 Q_q^2 \alpha \alpha_s~ \left\{ 1+ 
\frac{u-t}{s}\ln\left(\frac{u}{t}\right) + \frac{1}{2}\frac{t^2+u^2}{s^2} 
\left[\ln^2\left(\frac{u}{t}\right) + \pi^2\right]\right\} \; ; 
\label{pppp_sm} \\
{\cal M}^{++--}_{sm}(s,t,u) &=& i~ \delta_{ab}~ 8 Q_q^2 \alpha \alpha_s \; ; 
\label{ppmm_sm} \\
{\cal M}^{++-+}_{sm}(s,t,u) &=& -i~ \delta_{ab}~ 8 Q_q^2 \alpha \alpha_s\; ; 
\label{ppmp_sm}
\end{eqnarray}
with $\alpha_s$ being the strong coupling constant, $Q_q$ the quark charge,
$a$ and $b$ standing for the gluon colors, and $s$, $t$ and $u$ the Mandelstam
invariants.  The remaining helicity amplitudes can be obtained from the above
expressions through parity and crossing relations \cite{jikia}:
\begin{eqnarray}
{\cal M}^{\pm\pm\mp\pm}(s,t,u) &=& {\cal M}^{\pm\mp\pm\pm}(s,t,u) = 
{\cal M}^{\pm\mp\mp\mp}(s,t,u) = {\cal M}^{\pm\pm\pm\mp}(s,t,u) \; ; 
\label{cross1} \\
{\cal M}^{--++}(s,t,u) &=& {\cal M}^{++--}(s,t,u)\; ; 
\label{cross2} \\
{\cal M}^{\pm\mp\pm\mp}(s,t,u) &=& {\cal M}^{----}(u,t,s) = 
{\cal M}^{++++}(u,t,s)\; ; 
\label{cross3} \\
{\cal M}^{\pm\mp\mp\pm}(s,t,u) &=& {\cal M}^{++++}(t,u,s)\; .  
\label{cross4}
\end{eqnarray}

On the other hand, the top contribution is given by
\begin{eqnarray}
{\cal M}^{++++}_{sm}(s,t,u) &=& \frac{i}{2} \delta_{ab} 8Q_q^2 \alpha \alpha_s 
\left\{-2 + \frac{1}{s^2}\left[2(u^2-t^2)(B_0(t)-B_0(u)) + 
(C_0(t)t + C_0(u)u)\times\right.\right. 
\nonumber \\
&\times& (8m_t^2 s - 2t^2 - 2u^2)
+ (D_0(s,t)+D_0(s,u))s^2m_t^2(-4m_t^2+2s)+ \nonumber \\
&+& \left.\left.D_0(t,u)\left(-4m_t^4s^2 + 2m_t^2st^2 - 4m_t^2stu 
+ t^3u + 2m_t^2su^2 + tu^3\right)\right]\right\}\;\;;
\label{pppp_top} \\
{\cal M}^{++--}_{sm}(s,t,u) &=& i \delta_{ab} 8 Q_q^2 \alpha \alpha_s\left[1 - 
2m_t^4 \left(D_0(s,t) + D_0(s,u) + D_O(t,u)\right)\right]\;\;;
\label{ppmm_top} \\
{\cal M}^{++-+}_{sm}(s,t,u) &=& i \delta_{ab} 8 Q_q^2 \alpha \alpha_s 
\left\{-1 + 
\frac{1}{stu}\left[2m_t^2\left(C_0(s)s + C_0(t)t +C_0(u)u\right)\left(t^2+tu+
u^2\right)+ \right.\right. \nonumber \\
&+& 2m_t^4stu\left(D_0(s,t)+D_0(s,u)+D_0(t,u)\right) + 
m_t^2s^2t^2D_0(s,t) + \nonumber\\
&+& \left.\left. m_t^2s^2u^2D_0(s,u) + m_t^2t^2u^2D_0(t,u)\right]
\right\}\;\;, 
\label{ppmp_top}
\end{eqnarray}
where $m_t$ is the top quark mass and $B_0$, $C_0$ and $D_0$ are 
Passarino--Veltman functions \cite{pv} 
\begin{eqnarray}
B_0(s) &\equiv& B_0(s,m_t,m_t)\;\;; \label{b0} \\
C_0(s) &\equiv& C_0(0,0,s,m_t,m_t,m_t)\;\;; \label{c0} \\
D_0(s,t) &\equiv& D_0(0,0,0,0,s,t,m_t,m_t,m_t,m_t)\;\;. \label{d0}
\end{eqnarray}



\begin{table}
\begin{center}
\begin{tabular}{|c|ccccc|}
$n$ & 3 & 4 & 5 & 6 & 7 \\ \hline
$M_S$ (TeV)& 1.92  & 1.73  & 1.61  & 1.52 & 1.45 \\ 
\end{tabular}
\vskip 12pt
\caption{$2\sigma$ limits in TeV for the quantum gravity scale $M_S$ as a
function of the number of extra dimensions for the Tevatron Run
II.}
\label{limtab1}
\end{center}
\end{table}


\begin{table}
\begin{center}
\begin{tabular}{|c|ccccc|}
$n$ & 3 & 4 & 5 & 6 & 7 \\ \hline
${\cal L} = 10$ fb$^{-1}$ & 6.70 & 6.15 & 5.78 & 5.50 & 5.28 \\
${\cal L} = 100$ fb$^{-1}$& 8.50 & 7.70 & 7.16 & 6.79 & 6.50 \\ 
\end{tabular}
\vskip 12pt
\caption{$2\sigma$ limits in TeV for the extra dimensions gravity scale
$M_S$ as a function of the number of extra dimensions for the LHC with
luminosities ${\cal L} = 10$ and $100$ fb$^{-1}$.}
\label{limtab2}
\end{center}
\end{table}


\begin{figure}
\vskip 36pt
\begin{center}
\centerline{
\begin{picture}(450,80)(-15,-15)
\ArrowLine(0,20)(40,20)
\ArrowLine(40,60)(0,60)
\ArrowLine(40,20)(40,60)
\Photon(40,20)(80,20){4}{3}
\Photon(40,60)(80,60){4}{3}
\Text(20,10)[c]{$q$}
\Text(20,70)[c]{$\bar{q}$}
\Text(35,40)[r]{$q$}
\Text(85,20)[l]{$\gamma (k')$}
\Text(85,60)[l]{$\gamma (k)$}
\ArrowLine(150,20)(190,20)
\ArrowLine(190,60)(150,60)
\ArrowLine(190,20)(190,60)
\Photon(190,20)(230,60){4}{4}
\Photon(190,60)(230,20){4}{4}
\Text(170,10)[c]{$q$}
\Text(170,70)[c]{$\bar{q}$}
\Text(185,40)[r]{$q$}
\Text(235,20)[l]{$\gamma (k')$}
\Text(235,60)[l]{$\gamma (k)$}
\ArrowLine(300,20)(330,40)
\ArrowLine(330,40)(300,60)
\Photon(330,40)(360,40){4}{5}
\Photon(360,40)(330,40){-4}{5}
\Photon(360,40)(390,20){4}{3}
\Photon(360,40)(390,60){-4}{3}
\Text(300,20)[r]{$q$}
\Text(300,60)[r]{$\bar{q}$}
\Text(345,50)[c]{$KK$}
\Text(395,20)[l]{$\gamma (k')$}
\Text(395,60)[l]{$\gamma (k)$}
\end{picture}
}
\end{center}
\vskip -0.8cm
\caption{Feynman diagrams contributing to the subprocess $q\bar{q} \to 
\gamma\gamma$, including the Kaluza--Klein graviton exchange.}
\label{graph_qq}
\end{figure}
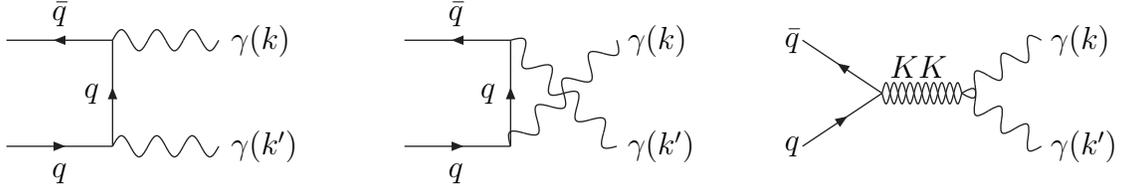


\newpage 
 

\begin{references}

\bibitem{gsw} See, for instance, M. B. Green, J. H. Schwarz, and
E. Witten, {\em Superstring Theory}, Cambridge Press (1987).

\bibitem{led} N. Arkani--Hamed, S. Dimopoulos, and G. Dvali,
Phys. Lett. {\bf B429}, 263 (1998); I. Antoniadis, N. Arkani-Hamed,
S. Dimopoulos, and G. Dvali, Phys. Lett. {\bf B436}, 257 (1998);
I. Antoniadis and C. Bachas, Phys. Lett. {\bf B450}, 83 (1999).

\bibitem{ph-astro}  N. Arkani--Hamed, S. Dimopoulos, and G. Dvali,
Phys. Rev. {\bf D59}, 086004 (1999).

\bibitem{astro} S.~Cullen and M.~Perelstein, hep-ph/9903422;
V. Barger, T. Han, C. Kao, and R. Zhang,  hep-ph/9905474.

\bibitem{KK} See, {\it e. g.}, {\it Modern Kaluza-Klein Theories}, eds. 
T. Appelquist, A. Chodos and P. Freund, Addison-Wesley Pub. (1987).

\bibitem{string0} 
E. Witten, Nucl. Phys. {\bf B443}, 85 (1995);
P. Ho\v{r}ava and E. Witten, Nucl. Phys. {\bf B460},
506 (1996); Nucl. Phys. {\bf B475}, 94 (1996).

\bibitem{string} For recent discussions, see for example,
C. Efthimiou and B. Greene (Ed.), {\it TASI96: Fields, Strings
and Duality}, World Scientific, Singapore (1997) and references therein.

\bibitem{scs} I. Antoniadis, Phys.~Lett.~{\bf B246}, 377 (1990); 
J. Lykken, Phys. Rev. {\bf D54}, 3693 (1996); 
K. Dienes, E. Dudas, and T. Ghergetta, Phys. Lett. {\bf B436}, 55 (1998).
G. Shiu and S.-H. H. Tye, Phys. Rev. {\bf D58}, 106007 (1998);
Z. Kakushadze and S.-H. H. Tye, Nucl. Phys. {\bf B548}, 180 (1999);
L. E. Ib\'a\~nez, C. Mu\~noz and S. Rigolin, hep-ph/9812397.

\bibitem{joanne} J. Hewett, Phys. Rev. Lett. {\bf 82}, 4765 (1999).

\bibitem{giudice} G. F. Giudice, R. Rattazzi and J. D. Wells, 
Nucl. Phys. {\bf B544}, 3 (1999); E. A. Mirabelli, M. Perelstein and M. E. 
Peskin, Phys. Rev. Lett. {\bf 82}, 2236 (1999).

\bibitem{mathews} P. Mathews, S. Raychaudhuri, and K. Sridhar, Phys. 
Lett. {\bf B450}, 343 (1999); T. Rizzo, hep-ph/9902273.

\bibitem{tao_grav} T. Han, J. Lykken, and Ren--Jie Zhang, 
Phys. Rev. {\bf D59}, 105006 (1999).

\bibitem{pheno} A. K. Gupta, N. K. Mondal, and S. Raychaudhuri,
hep-ph/9904234; P. Mathews, S. Raychaudhuri, and K. Sridhar, Phys. Lett. {\bf
B455}, 115 (1999) and hep-ph/9904232; T. Rizzo, Phys. Rev. {\bf D59}, 115010
(1999), hep-ph/9903475 and hep-ph/9904380; S.Y. Choi {\it et al.},
Phys. Rev. {\bf D60}, 013007 (1999); K. Agashe and N.G. Deshpande,
Phys. Lett. {\bf B456}, 60 (1999); K. Cheung and W.-Y. Keung, hep-ph/9903294;
D. Atwood, S. Bar-Shalom, and A. Soni, hep-ph/9903538; C. Balazs {\it et al.},
hep-ph/9904220; G. Shiu, R. Shrock, and S.-H. H. Tye, hep-ph/9904262;
K. Y. Lee, H.S. Song, and J. Song, hep-ph/9904355; Hooman Davoudiasl,
hep-ph/9904425; K. Cheung, hep-ph/9904510; T. Han, D. Rainwater, and
D. Zeppenfeld, hep-ph/9905423; Xiao-Gang He, hep-ph/9905500.

\bibitem{kingman} K. Cheung, hep-ph/9904266.

\bibitem{old:gg} V. Constantini, B. de Tollis, and G. Pistoni, Nuovo Cimento
{\bf 2A}, 733 (1971); Ll. Ametler, {\em et al.}, Phys. Rev. {\bf D32}, 1699
(1985); D. Dicus and S Willenbrock, Phys. Rev. {\bf D37}, 1801 (1988).

\bibitem{pdg} Particle Data Group, Eur. Phys. J. {\bf C3}, 1 (1998).

\bibitem{cdf-d0} F. Abe {\em et al.}, CDF Collaboration, Phys. Rev. D{\bf 59},
092002 (1999); {\it ``Searching for Higgs Mass Photon Pair in} $p\bar{p}$ {\it
Collitions at} $\protect\sqrt{s} = 1.8$ {\it TeV''}, submitted by P. Wilson
(CDF Collaboration) to ICHEP'98; {\it ``Direct Photon Measurements at D0''},
submitted by D0 Collaboration to ICHEP'98.

\bibitem{mrsg} A. D. Martin, R. G. Roberts, and W. J. Stirling, Phys. Lett. 
{\bf B354}, 155 (1995).

\bibitem{outros} D. Atwood, S. Bar-Shalom, and A. Soni, hep-ph/9906400.
 
\bibitem{jikia} G. Jikia and A. Tkabladze, Phys. Lett. {\bf B323}, 453 (1994).

\bibitem{pv} G. Passarino and M. Veltman, Nucl. Phys. {\bf B160}, 151 (1979).



\end{references}
\end{document}